\documentclass[12pt]{article}
\usepackage{epsfig}
\usepackage{amsmath}
\usepackage{amsfonts}

\begin{document}
\pagestyle{empty}
\begin{flushright}
UMN-TH-2524/06\\
October 2006
\end{flushright}
\vspace*{5mm}

\begin{center}
{\Large\bf Holography with Schr\"odinger Potentials}
\vspace{1.0cm}

{\sc Brian Batell$^a$}\footnote{E-mail:  batell@physics.umn.edu}
{\small  and}
{\sc Andrew Larkoski$^b$}\footnote{E-mail: larkoa@u.washington.edu}
\\
\vspace{.5cm}
{\it\small {$^a$School of Physics and Astronomy\\
University of Minnesota\\
Minneapolis, MN 55455, USA}}\\
\vspace{.5cm}
{\it\small {$^b$Department of Physics\\
University of Washington\\
Seattle, WA 98195, USA}}\\
\end{center}

\vspace{1cm}
\begin{abstract}

We examine the analogue one-dimensional quantum mechanics problem associated with bulk scalars and fermions in a slice of AdS$_5$. The ``Schr\"odinger'' potential can take on different qualitative shapes depending on the values of the mass parameters in the bulk theory. Several interesting correlations between the shape of the Schr\"odinger potential and the holographic theory exist. We show that the quantum mechanical picture is a useful guide to the holographic theory by examining applications from phenomenology. 

\end{abstract}

\vfill
\begin{flushleft}
\end{flushleft}
\eject
\pagestyle{empty}
\setcounter{page}{1}
\setcounter{footnote}{0}
\pagestyle{plain}

\section{Introduction}

The Randall-Sundrum (RS) model provides an attractive explanation for the large hierarchy that exists between the Planck and weak scales\cite{rs}. In this five-dimensional (5D) model, an Ultraviolet (UV) and Infrared (IR) brane are placed at orbifold fixed points, providing boundaries to the bulk Anti-de Sitter (AdS) geometry.  Due to the ``warped'' background geometry of Anti-de Sitter space, energy scales depend on location in the extra dimension, and the hierarchy is simply a consequence of placing the standard model, and in particular the Higgs boson, on the IR brane. 

In fact, the hierarchy can also be understood by recasting the problem into an analogue 1D quantum mechanics problem, as done originally in \cite{rs2}. In this picture, the graviton is subject to a ``volcano'' potential. The massless mode, which is the 4D massless graviton, is bound in the delta-function ($\delta$) well located at the position of the UV brane, and thus overlaps very weakly with the IR brane. Thus, the gravitational interaction with the standard model fields on the IR brane is feeble. 

An alternative interpretation of the hierarchy in this framework can be motivated by the AdS/CFT correspondence \cite{adscft1,adscft2,adscft3, pheno1, pheno2, pheno3}. The 5D gravitational theory is postulated to be dual to a strongly coupled 4D conformal field theory (CFT). In this interpretation, the Planck scale is the UV cutoff for the CFT. Between the the Planck and weak scales, the theory is conformal. As we move below the weak scale, conformal symmetry is broken, and the CFT produces the composite standard model fields. Although the standard model is composite, the massless 4D graviton is associated with the dynamical elementary source field.

Over the past few years a ``dictionary'' has been created that relates a variety of bulk theories in a slice of AdS$_5$ to their dual 4D counterpart theories. This connection has been elaborated on in more general theories with bulk scalars\cite{pheno2, pheno3, lectures}, fermions\cite{ferm}, gauge fields\cite{pheno1, ad, u1}, and gravitons\cite{pheno1, pheno2, pheno3, emerge}, and the holographic description can have different qualitative features depending on the parameters in the bulk theory. In particular, depending on the value of the mass parameters in the 5D theory, the dual description will be described by a particular branch in which the massless mode is either primarily composed of the elementary source field or is a CFT bound state.

In this paper, we examine the analogue 1D quantum mechanics problem in these more general theories. As in the original RS models, the ``Schr\"odinger'' potential provides insight to the localization properties and masses of modes in the 5D theory. Furthermore, the potential is consistent with the dual description of these models, and we find a number of interesting correlations between the shape of the Schr\"odinger potential and the branch structure of the dual theory. In particular, the potential assumes its minimum value precisely at the point where the dual interpretation changes, and there exists a symmetry about this transition point in the bulk potential which mimics a symmetry in the scaling dimension of the dual CFT operator. Therefore, the quantum mechanical picture provides a useful guide to the nature of the dual theory. Our analysis suggests that perhaps other Schr\"odinger potentials with similar localization properties may admit a dual holographic interpretation.

We lay the groundwork in Section 2 by presenting, in the language of 1D quantum mechanics, a simple example of a scalar field propagating in flat space subject to boundary delta-function potentials. In Sections 3 and 4, we analyze the Schr\"odinger potential for bulk scalars and fermions living in a slice of AdS$_5$, and make the connection to the holographic interpretation. In Section 5 we use the Schr\"odinger picture to examine applications from phenomenology, and finally, we present our conclusions in Section 6.

\section{Localization in the fifth dimension} 

Theories with compact extra dimensions utilize mathematics familiar from elementary electromagnetism and quantum mechanics. In particular, motion in the fifth dimension of 5D theories can be reformulated into an equivalent 1D quantum mechanics problem. A familiar problem from 1D quantum mechanics is a particle under the influence of a delta-function potential. In this section we will show how this problem finds an analogue in 5D theories containing localized massless modes.

Consider a real massive scalar field in 5D Minkowski space $(x^\mu,y)$, with metric $\eta_{\mu\nu}={\rm diag}(-,+,+,+,+)$. The fifth dimension $y$ is compactified on a $Z_2$ orbifold with a UV (IR) brane at the orbifold fixed points $y=0 (\pi R)$.  The 5D action is:
\begin{equation}
S= -\int d^4x~dy \Big[\frac{1}{2} (\partial_M \Phi)^2+\frac{1}{2} m^2\Phi^2 + \mu\Phi^2\left(\delta(y)-\delta(y-\pi R)\right)\Big],\label{action}
\end{equation}
where the last term is a scalar mass localized on each brane. We will see shortly that the boundary mass parameter $\mu$ must take a specific value to permit a massless zero mode solution consistent with the boundary conditions.

To find the equation of motion for $\Phi(x,y)$ we perform a Kaluza-Klein decomposition by expanding the field $\Phi$ in a complete set of states:
$\Phi(x,y)=\sum_{n=0}^\infty \phi_n(x) f_n(y)$,
where $\phi_n(x)$ satisfies $\Box \phi_n(x)=m_n^2\phi_n(x)$ and the eigenfunctions satisfy the orthonormality condition
$\int_0^{\pi R} dy f_n(y)f_m(y)=\delta_{nm}$.
The equation of motion for $f_n(y)$ becomes
\begin{equation}
\Big[\partial_5^2-m^2 -2\mu(\delta(y)-\delta(y-\pi R))\Big]f_n(y)=-m_n^2 f_n(y)~. \label{s1}
\end{equation}

Notice that the bulk equations of motion do in fact allow a zero mode solution, which is given by
\begin{equation}
f_0(y)= A e^{m |y|} + B e^{- m|y|}
 \label{0mode1},
\end{equation}
It is customary to impose either Dirichlet ($\Phi|_{0,\pi R} =0$) or Neumann ($\partial_5 \Phi|_{0,\pi R} =0$) boundary conditions. The zero mode solution is not compatible with these boundary conditions, and thus, only massive modes exist. The purpose of adding the brane-localized mass term in (\ref{action}) is to impose a boundary condition consistent with having a nontrivial zero mode solution. The boundary condition can be derived either by varying the action or by integrating the equation of motion and is given by
\begin{equation}
(\partial_5 -\mu) \Phi(x,y)\bigg\vert_{0,\pi R}=0~.
 \label{bc0}
\end{equation}
Inserting the zero mode solution (\ref{0mode1}) into (\ref{bc0}), we can see that the boundary mass parameter must be
$\mu=\pm m$
in order to have either $A$ or $B$ nonvanishing. If this is the case, the theory contains a massless scalar field in the effective theory. The normalized zero mode solution is then given by
\begin{equation}
f_0(y) = \sqrt{\frac{ \pm 2m}{e^{\pm 2 m \pi R}-1}} e^{\pm m |y|}=\sqrt{\frac{ 2\mu}{e^{ 2 \mu \pi R}-1}} e^{\mu |y|},\label{0mode2}
\end{equation}
where we have written the solution both in terms of the bulk and boundary mass parameters. The boundary mass parameter can take on any real value, so that the zero mode can be localized anywhere in the bulk. 

Let us now consider the massive modes for the choice $\mu=\pm m$, needed to have a massless mode.  When $m_n^2<m^2$ there are no solutions consistent with the boundary condition (\ref{bc0}). However for $m_n^2>m^2$ the normalized massive mode solutions are
\begin{equation}
f_n(y)= \sqrt{\frac{2}{\pi R}} \frac{\mu}{m_n}\left( \sin{\frac{n|y|}{R}} + \frac{n}{\mu R} \cos{\frac{ny}{R}} \right), \label{massmode1} 
\end{equation}
where the mass of the $n$th mode is given by
$m_n^2 = m^2 +\left(\frac{n}{R}\right)^2$ for  $n=1,2,\dots$,
as typically is the case when compactifying 5D theories in flat space.

\subsection{Quantum mechanical analogue}

\begin{figure}
\centerline{
\includegraphics[width=1\textwidth]{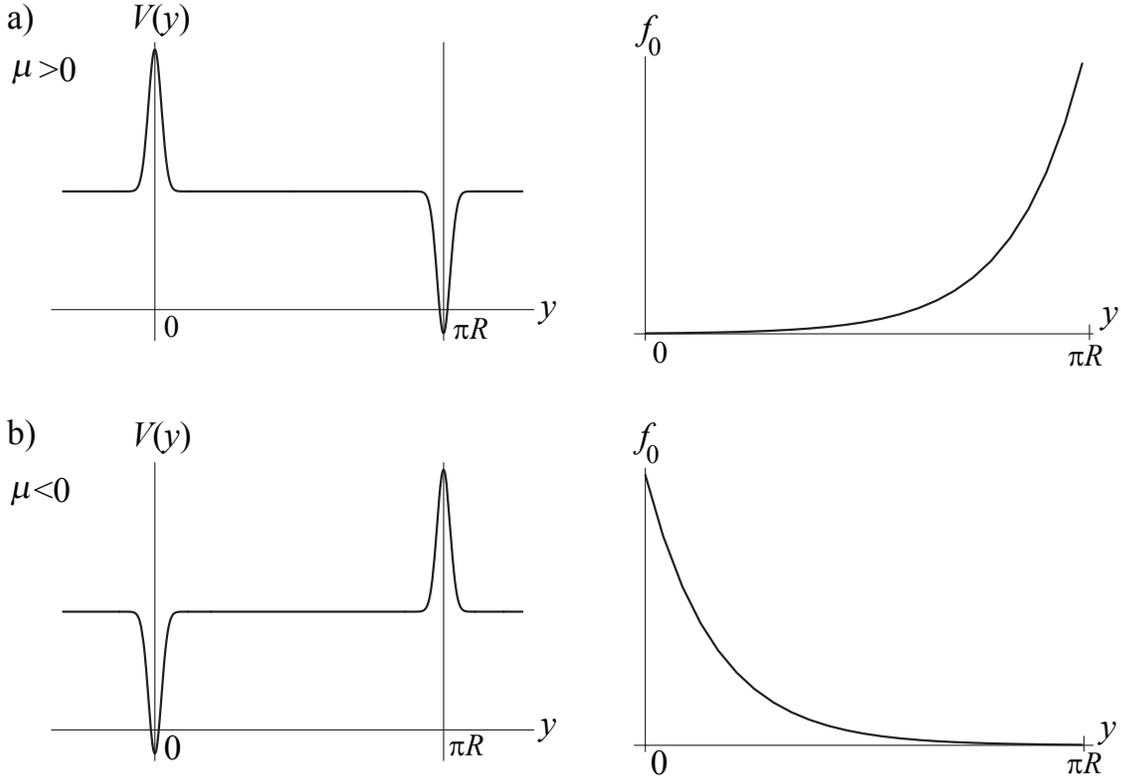}}
\caption{The Schr\"odinger potential $V(y)$ and zero mode wave function $f_0$ in the flat extra dimension.}
\label{fig1}
\end{figure}

It is illuminating to describe this 5D problem in the language of one-dimensional quantum mechanics. The localization features of the zero modes as well as the behavior of the massive modes can easily be understood from this point of view. The equation of motion (\ref{s1}) can be interpreted as  a nonrelativistic  ``Schr\"odinger'' equation, $[p^2/2 +V(y)]\psi(y)=E\psi(y)$, where the derivative $\partial_5$ is the ``momentum", and the masses $m_n$ are the ``energy" eigenvalues. We can then read off the potential from (\ref{s1}):
\begin{equation}
V(y) = \frac{1}{2}m^2 \pm m(\delta(y)-\delta(y-\pi R))=\frac{1}{2}\mu^2 +\mu(\delta(y)-\delta(y-\pi R)),
\end{equation}
which is shown in Fig \ref{fig1}. The bulk mass simply corresponds to a constant potential in the $y$-direction, while at the boundaries there exists either a delta-function ($\delta$) well or barrier with strength $m$. It is simpler to characterize the potential in terms of the boundary mass parameter $\mu=\pm m$, which can continuously range from $-\infty<\mu<\infty$. 

We see that in the bulk, the potential has a minimum at $\mu=m=0$, at which point the delta-functions also turn off. The zero mode has a flat profile in this case, $f_0(y)=\sqrt{1/\pi R}$, which can be seen by taking the limit $m \rightarrow 0$ in (\ref{0mode2}).  When the boundary mass parameter is positive, 
$\mu>0$, the $\delta$ barrier (well) is at $y=0 (\pi R)$, as depicted in Fig. \ref{fig1}a. From quantum mechanics, we know that a delta-function well supports a single bound state when boundary conditions are imposed at infinity. In this case, we are working with a finite dimension so that all states are bound states, but we can easily see from (\ref{0mode2}) that the zero mode scalar field is localized near $y=\pi R$, and therefore corresponds to the bound state supported by the $\delta$ well. The strength of the well $\mu$ reflects the degree of localization of the zero mode, as indicated in (\ref{0mode2}). All of these features are familiar from the one-dimensional delta-function potential in quantum mechanics~\cite{griffiths}. Finally if $\mu<0$, the localization is reversed as shown in Fig. \ref{fig1}b. The well (barrier) is located at $y=0 (\pi R)$, and the zero mode is localized near $y=0$ as it should be.

The behavior of the massive modes can also be understood from a quantum mechanics perspective. Since the derivative $\partial_5$ corresponds to the momentum $p_n$, we see from the massive wavefunctions (\ref{massmode1}) that $p^2_n \sim n^2$. In the quantum mechanics picture the higher modes have a larger momentum and are therefore influenced less by the potential than the low modes. 
Also since the potential is a constant the massive modes are not localized near any of the boundaries. 
In particular these modes are not normalizable when the UV boundary is removed. This is a crucial
requirement for the holographic description.

Using our intuition from simple 1D quantum mechanics problems is very helpful when trying to understand the properties of Kaluza-Klein modes in theories with extra-dimensions. We have warmed up with the flat space example where delta-function wells (brane-localized mass terms) force zero modes to be localized in the extra dimension. A similar analysis is helpful for warped models, and in fact the physics is much more interesting because of the connection with holography. Many features of the holographic theory can be identified by examining the corresponding Schr\"odinger potential.

\section{Scalar holography}

We first analyze the simplest case, that of a scalar field in the bulk of AdS$_5$. Our goal in this section is to analyze the general Schr\"odinger potential required to localize massless modes anywhere in the warped bulk and make the connection between the quantum mechanics picture and the dual interpretation of the 5D theory as a broken conformal field theory.

The metric in a slice of AdS$_5$ is
\begin{equation}
ds^2=A^2(z)(\eta_{\mu\nu}dx^{\mu}dx^{\nu}+dz^2)~,
\label{metric}
\end{equation}
where the warp factor is $A(z)=(k|z|+1)^{-1}$ with $k$ the AdS curvature scale. We will denote 5D indices with Latin letters $(A,B,...)$ and 4D Lorentz indices with Greek letters ($\mu,\nu, ...)$. This is the Randall-Sundrum solution (RS1) with two opposite tension 3-branes and a bulk cosmological constant~\cite{rs}. In RS1 the UV and IR branes are located at the orbifold fixed points $z=0$ and $z=z_*$, respectively. It is often useful for holography to consider limits in which the locations of the branes differ from these fixed points.     

Ref. \cite{lectures} analyzes in detail the problem of localizing massless scalar modes at any position in the extra dimension and gives the detailed dual description of the theory. Like the flat case (\ref{action}), we must add brane-localized mass terms to the bulk action to localize a massless mode. The action is 
\begin{equation}
S=-\int d^4x dz \; \sqrt{-g}\;\Big[\frac{1}{2}(\partial_{M}\Phi)^2+\frac{1}{2}ak^2\Phi^2 + b k A^{-1}(z) 
\left(\delta(z)-\delta(z-z_*)\right)\Phi^2\Big]~,
\label{actions}
\end{equation} 
where the bulk and boundary masses have been expressed in terms of the bulk curvature $k$ with dimensionless parameters $a$ and $b$. To localize a massless mode, the parameters must be related as $b=2\pm\sqrt{4+a}$. 

This 5D gravity description can be given a purely 4D interpretation via the AdS/CFT 
correspondence~\cite{adscft1,adscft2,adscft3}, which relates type IIB string theory compactified on AdS$_5\times$ S$^5$ and ${\cal N}=4$ supersymmetric gauge theory. 
Motivated by the string correspondence, weakly coupled phenomenological theories in a slice of AdS$_5$ can be given a holographic description as a strongly coupled (broken) 4D conformal field theory interacting with a dynamical elementary source field~\cite{pheno1, pheno2, pheno3}. 
The holographic procedure is straightforward as reviewed in Ref.~\cite{lectures}. For every bulk field $\Phi(x,z)$ in the gravity theory, a corresponding operator ${\cal O}$ exists in the CFT. Arbitrary boundary conditions are applied to the field in the UV, $\Phi(x,z=0)=\varphi^0$, and the effective action is computed. This boundary value $\varphi^0$ is interpreted as a source in the partition function for the operator ${\cal O}$. Correlation functions containing products of ${\cal O}$ can then be directly computed from the 5D side by taking functional derivatives of the effective action. 

There are two branches characterized by the value of the boundary mass parameter $b$ which have different dual descriptions: the $b_-$ branch for $b<2$, in which the massless particle is mostly elementary source, and the $b_+$ branch for $b>2$ where the massless particle is a composite CFT state. In the region near $b=2$, there is strong mixing between the source and CFT fields. We can compute the two point function for the operator ${\cal O}$ using the holographic procedure~\cite{lectures}
\begin{equation}
  \langle {\cal O}(p){\cal O}(-p)\rangle =\Sigma(p) = \frac{k}{g_{\varphi}^2}
     \frac{q_0 (I_{b-1}(q_0)K_{b-1}(q_*)-I_{b-1}(q_*)K_{b-1}(q_0))}
     {I_{b-2}(q_0)K_{b-1}(q_*)+I_{b-1}(q_*)K_{b-2}(q_0)}~,
\end{equation}
where we have included the scalar coupling $g_{\varphi}$ in front of the correlator, and defined the variable $q_i=p/(kA(z_i))$ related to the momentum.

To interpret the dual theory, we expand the two point function in powers of momentum above and below the infrared scale. Here we briefly summarize the results: On the $b_-$ branch, the leading analytic term is proportional to $p^2$. The massless state is primarily the source field on this branch. On the $b_+$ branch, the expansion also yields a constant term, which indicates the source is massive on this branch. When we expand the correlator at low energy on the $b_+$ branch, we find a pole at $p=0$ indicating that the massless particle in the dual theory is a CFT bound state. We will see this in detail in the next section and find a correspondence between the qualitative features of the dual theory and the shape of the Schr\"odinger potential.

\subsection{Schr\"odinger potential}

We can make an entry into the ``dictionary'' by relating the shape of the Schr\"odinger potential and the elementarity/compositeness of the massless mode. As in the case of RS gravity~\cite{rs2}, when there is a ``volcano'' potential (Fig. \ref{fig2}a), the zero mass eigenstate will be primarily a source field in the dual description. However, the Schr\"odinger potential can have other shapes. For example, rather than a volcano there can exist a shallow bulk well near the UV brane and a delta-function ($\delta$) well at $z=z_*$ (Fig. \ref{fig2}c).  In this case we will see that there is strong mixing between the CFT and source fields. Another possibility is that we can have a ``plugged volcano'': the bulk potential has the shape of a volcano, but there is a $\delta$ barrier at the origin rather than a $\delta$ well (Fig. \ref{fig2}e). In this case, there is a $\delta$ well at the IR brane, corresponding to a massless CFT bound state in the boundary theory. Just as examining the localization features of the modes in the 5D theory can give a quick qualitative picture of the 4D description, the shape of the quantum mechanical potential can yield insight into the nature of the dual theory. Indeed, both methods are complementary, and of course, the shape of the potential can quickly give us information regarding the localization of modes.  In fact our method can be generalized to conjecture that any Schr\"odinger potential with similar localization properties admits a possible dual holographic interpretation, and to suggest this we will present our results in terms of the warp factor $A(z)$.

With this motivation, let us write the equation of motion in the form of a 1D time-independent Schr\"odinger equation: $[p^2/2 +V(z)]\psi(z)=E\psi(z)$. Expanding the field 
\begin{equation}
\Phi(x,z)=\sum_{n=0}^{\infty} \phi_n(x) A^{-3/2}(z) g_n(z),
\end{equation}
and using the relation $b=2\pm\sqrt{4+a}$, the equation of motion for $g_n(z)$ is 
\begin{equation}
\left(-\frac{1}{2}\partial^2_z+V(z) \right)g_n(z)=\frac{m_n^2}{2}g_n(z)~,
\label{schr1}
\end{equation}
where the Schr\"odinger potential is defined as
\begin{equation}
V(z)=\frac{1}{2}\left((b-2)^2-\frac{1}{4}\right) k^2 A^2(z)+\left(b-\frac{3}{2}\right)k A(z) \left[\delta(z)-\delta(z-z_*)\right]~.
\label{pot1}
\end{equation}
The RS limit is obtained for $b=0$ and agrees with the result in Ref~\cite{rs2}.
The solutions $g_n(z)$ are given by
\begin{equation}
g_n(z)= N_n A^{-1/2}(z) \biggl[J_{\alpha}\biggl(\frac{m_n}{k A(z)}\biggr)+\beta_n Y_{\alpha}\biggl(\frac{m_n}{k A(z)} \biggr)\biggr],
\label{sol1}
\end{equation}
where $\alpha =\sqrt{4+a}$ and $N_n, \beta_n$ are constants determined from the boundary conditions and normalization.  
Note that these are related to the physical propagating modes $f_n$ by  $f_n(z)=A^{-3/2}(z) g_n(z)$.
In particular $g_0(z) \propto A^{\mp |b-3/2|}(z)$ for $b>3/2 (b<3/2)$, which is consistent with the strength of the delta-function potential in $V(z)$.

The properties of the Schr\"odinger potential illuminate many features of the Kaluza-Klein modes and can be compared with the holographic interpretation of the 5D theory. The shape of the potential is determined solely by the boundary mass parameter $b$, which of course also determines the wavefunctions of the KK modes. We plot $V(z)$ in Fig. \ref{fig2} for various values of $b$. Notice that the bulk potential is positive for $b<3/2$ and $b>5/2$ and takes the shape of a volcano at $z=0$. Between $3/2 < b<5/2$ the potential is negative and forms a well localized at the origin. At $b=3/2$, the potential is uniformly zero as the $\delta$ well and barrier vanish. Let us consider three separate regions of interest: $b<3/2$, $3/2<b<5/2$, and $b>5/2$, corresponding in the dual theory to an elementary, strongly mixed, and composite massless particle. In each case we will emphasize the connection with the holographic picture. 

\subsubsection{$b<3/2$}

\begin{figure}
\centerline{
\includegraphics[width=1\textwidth]{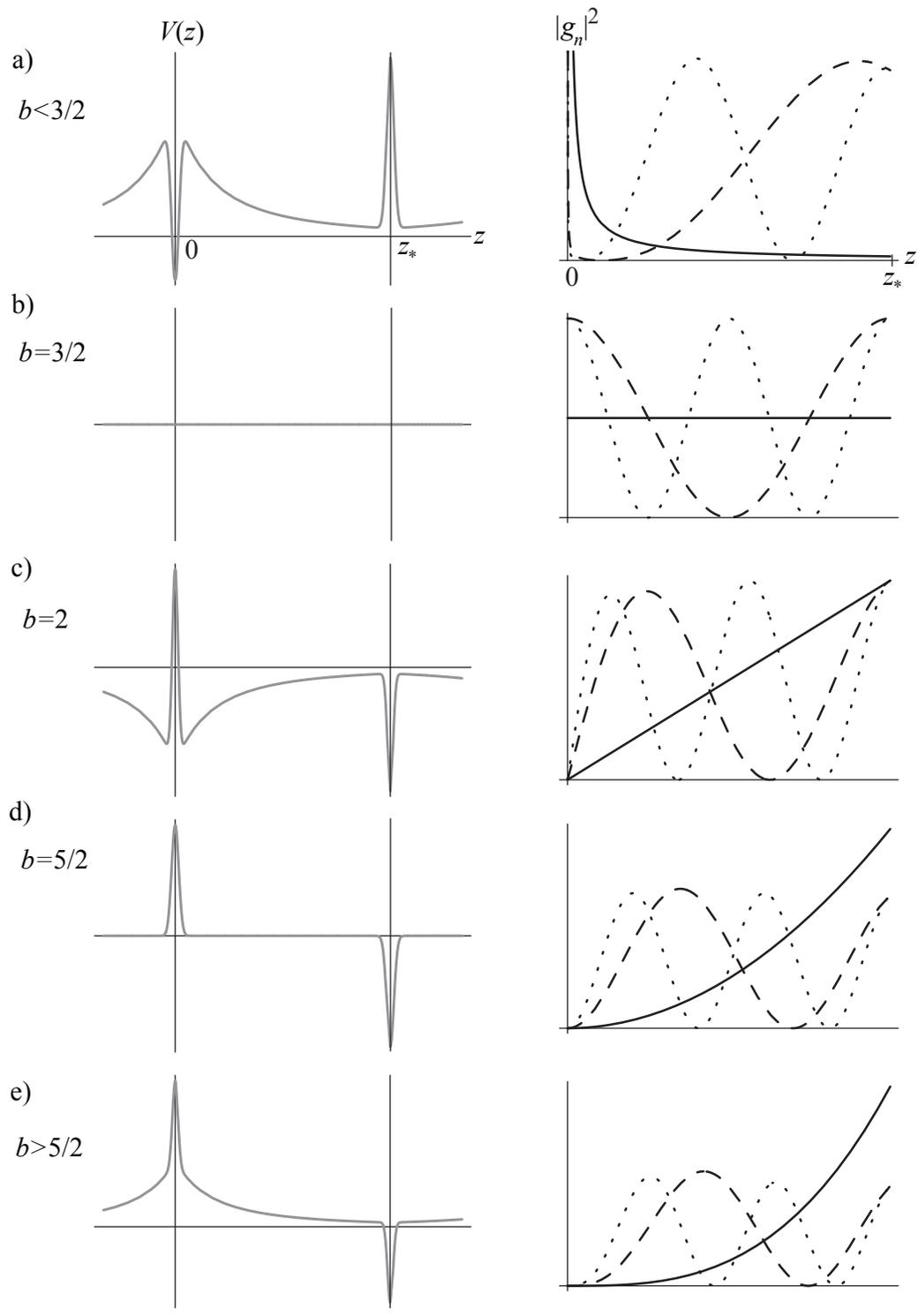}}
\caption{The Schr\"odinger potential $V(z)$ for the bulk scalar in warped space together with the
solution $|g_n|^2$. The solid, dashed, dotted line corresponds to the $n=0,1,2$ modes, respectively.}
\label{fig2}
\end{figure}

For $b<3/2$ the fields are subject to the volcano potential depicted in Fig. \ref{fig2}a. As in the case of RS gravity, the zero mode wavefunction $g_0$ is confined by the delta well located at $z=0$. Although the Kaluza-Klein tower is also influenced by the $\delta$ well, the modes are not confined and the wavefunctions leak out into the bulk. From a quantum mechanical picture this makes complete sense since a $\delta$ well can only support a single bound state.

Let us now examine the dual theory. We can expand the propagator for $b<3/2$ in the regime 
$q_1\gg 1$ (high energy)\cite{lectures}: 
\begin{equation}
\label{bmsig}
    \Sigma(p)\simeq -\frac{2k}{g_\varphi^2}\left[\frac{1}{1-b}
     \left(\frac{q_0}{2}\right)^2 + \left(\frac{q_0}{2}\right)^{4-2b} 
       \frac{\Gamma(b-1)}{\Gamma(2-b)}+\dots\right]~,
\end{equation}
Note that this expression is valid for noninteger $b$, while for integer $b$ a similar momentum expansion has log terms.

The nonanalytic term corresponds to the CFT correlator $ \langle {\cal O}{\cal O}\rangle_{CFT}$. Fourier transforming this term, we find the scaling dimension of the operator ${\cal O}$ to be dim ${\cal O}=4-b$. The leading analytic piece indicates that the source field propagates. Furthermore, we know it is massless by the absence of a constant analytic term. For $b<3/2$ the massless particle in the dual theory is primarily composed of the source field with a negligible contribution arising from the CFT fields. 

With this interpretation, at a scale $\Lambda \sim k$ we can write the dual theory as \cite{lectures}
\begin{equation}
\label{numL}
    {\cal L}_{4D} = -Z_0 (\partial\varphi^0)^2 + 
      \omega\Lambda^{b-1}~\varphi^0 {\cal O}+{\cal L}_{CFT}~,
\end{equation}
where the dimensionless couplings $Z_0$ and $\omega$ can be extracted from the two point function (\ref{bmsig}).
The elementary source field mixes with the CFT fields through its interaction with ${\cal O}$. The mass eigenstates will therefore be a mixture of source and CFT fields. The strength of this mixing depends on the value of $b$. Since the coupling is irrelevant for $b<1$, the massless particle will be primarily composed of the elementary source field. On the other hand, for $b>1$, the coupling has positive mass dimension and is therefore relevant. In this case, the mixing from the CFT cannot be neglected. 

As we approach $b=3/2$, the strength of the $\delta$ potentials, indicated by the coefficient ($b-3/2$), becomes weaker and weaker, and thus the zero mode becomes flatter. Exactly at $b=3/2$, the potential vanishes (in the bulk and on the boundaries). The massless mode has a flat profie, $g_0(z)=c$. The massive modes are purely sinusoidal and are (in some sense) delocalized in the bulk as can be seen in Fig.\ref{fig2}b. Note that although the fields are delocalized, the massive modes still couple more strongly at the fixed points than the zero modes.
 
At this point the behavior of the potential appears to be at variance with the known localization
properties of the massless mode. Indeed with respect to a flat metric the wavefunction of the massless mode is $\tilde{f_0}\sim e^{-ky }f_0\sim e^{(b-1)ky} $, which shows that the massless mode is  localized near the IR brane for $b>1$, where $(k|z|+1)=e^{k|y|}$. There is no conflict because the potential only describes the modes $g_n(z)$ which are not the same as the physical mode wavefunction $f_n= e^{3ky/2} g_n$. However the IR boundary coupling of the modes (either $f_n$ or $g_n$) suggests that for $0<b<2$, the massless state in the dual theory is primarily a source field (when we may have guessed it was composite). It appears that {\it only} considering the localization of modes as a guide to the dual theory leaves some ambiguity. What is important is the relative couplings of the modes to the IR brane, which can be seen regardless of our choice of coordinates. The reason this is the case will become clear in the next section.

Very roughly then, we can say that when the bulk field $\Phi$ is subject to the volcano potential ($b<3/2$) in Fig. \ref{fig2}, the dual description is an elementary (source) massless particle coupled to a strongly interacting CFT sector. Between $1<b<3/2$, the situation is more intricate; there is relevant mixing between the source and CFT fields, and therefore the mass eigenstates have sizable contributions from both sectors.

\subsubsection{$3/2<b<5/2$}

The Schr\"odinger potential becomes negative\footnote{Since the potential is negative in this region, it is conceivable that tachyonic modes may be allowed. However, one can show by direct calculation that no such modes exist.} in the region close to $b=2$, depicted in Fig. \ref{fig2}c. The volcano is flipped upside down, and the KK modes feel a shallow well formed by the $\delta$ barrier at $z=0$ and the negative $\sim 1/z^2$ bulk potential. Also notice that for $b>3/2$ the $\delta$ well is now located at $z=z_*$. From our intuition in quantum mechanics we know that the zero mode is the lone bound state of this $\delta$ well. Thus, as we increase $b$, the zero mode becomes more and more localized on the IR brane. 
 
Between $3/2<b<2$, the expansion of the correlator is the same as in (\ref{bmsig}) and thus the dual theory is still governed by the Lagrangian (\ref{numL}). For $b>2$ however, the expansion is different. The massless mode in the 5D theory becomes mostly a composite particle as can be seen by expanding the correlator below the IR momentum scale (for noninteger $b$)\cite{lectures}:
\begin{equation}
\label{IRcorr}
  \Sigma(p) \simeq -\frac{2k}{g_\varphi^2}\left[ (b-2) + 
   \left(\frac{q_0}{2}\right)^2 \frac{1}{(b-3)}  - (b-1)(b-2)^2 \; 
   \frac{A_*^{2b-2}}{A_0^{2b-2}}\,\left(\frac{2}{q_0}\right)^2\right]~.
\end{equation}
The source is still dynamical, but the leading constant term tells us that the source has become massive. More importantly, the correlator has produced a pole at $p=0$, indicating a massless bound state in the CFT spectrum.  As can be seen from the high energy expansion (\ref{UVcorr}) (for noninteger $b$)\cite{lectures},
\begin{equation}
\label{UVcorr}
 \Sigma(p)\simeq -\frac{2k}{g_\varphi^2}\left[ (b-2) +  \left(\frac{q_0}{2}
  \right)^2\frac{1}{(b-3)}+ \left(\frac{q_0}{2}\right)^{2b-4}
\frac{\Gamma(3-b)}{\Gamma(b-2)}\right]~.
\end{equation}
above the IR scale this pole disappears from the correlator and conformal symmetry is restored, truly telling us that this massless particle is composite.

Clearly $b=2$ is the crossover point in the dual interpretation of the massless mode; the composition of the massless particle in the 4D theory changes from mostly elementary source to mostly composite CFT state. This is also the point where the bulk mass $a=-4$ corresponding to the
Breitenlohner-Freedman bound for the stability of AdS space~\cite{bl}.
Back to the Schr\"odinger picture, at the transition point $b=2$, the potential $V(z)$ assumes its minimum value (see Fig. \ref{fig2}c). This statement is, in fact, quite general as we will see for the fermion case\footnote{We have also checked that this is the case for gauge bosons and gravitons with zero modes localized at different points in the bulk.}. Given a particular AdS$_5$ theory and its corresponding Schr\"odinger equation, {\it the minimum of the quantum mechanical potential corresponds to the transition point in the dual interpretation.}

At the crossover $b=2$, the potential well is at its deepest point. Qualitatively, we can say that the deeper the quantum mechanical well is, the more balanced are the contributions of the source and CFT fields to the massless eigenstate in the 4D theory. In other words, the strongest mixing occurs when the well is deepest. 

More light can be shed on the transition point by considering the IR couplings of the zero mode and the KK, as highlighted in Fig. \ref{fig2}c. For $b\ll2$, the KK modes couple much more strongly than the massless mode.  As we move to $b=2$, the zero mode couples with the same strength as the Kaluza-Klein modes. Past this point, for $b\gg2$, the zero mode couples much more strongly than the massive modes. Thus, as we emphasized previously, the dual description of the bulk theory does not depend entirely on where the zero mode is localized, but rather its relative localization compared to the massive modes. More clearly, the dual interpretation depends on the couplings of the modes to the IR branes.  

Let's think about the physics behind this observation from the viewpoint of holography. To be concrete, let us consider the case $b>2$, depicted in Fig. \ref{fig2}d and Fig. \ref{fig2}e. If we are living on the IR brane, then physics is dominated by the massless mode since it couples more strongly than the Kaluza Klein modes. Since location in the extra dimension corresponds to the momentum scale in the dual theory, the low energy expansion (\ref{IRcorr}) is the holographic equivalent of living on the IR brane. The pole in the correlator tells us about the dominance of the massless mode, while the source field is massive and is thus integrated out of the theory.  
As we move out into the bulk, the profile of the zero mode drops dramatically and the massive modes are more important. This corresponds to the high energy limit in the 4D theory; the pole disappears (the zero mode is composite) and the theory becomes conformal. On the other hand, for $b<2$ the zero mode still contributes strongly to bulk physics, and therefore in the dual theory, we find a (mostly) elementary massless state which exists above the IR scale.  

\subsubsection{$b>5/2$}

As we increase $b$ the potential well in the bulk becomes shallower until it finally vanishes at $b=5/2$. However at this point the brane potentials do not vanish.  The massive modes are once again purely sinusoidal as we would expect from our analysis of the flat space case in Section 1. The delta-functions allow a nontrivial zero mode solution. This situation is shown in Fig.\ref{fig2}d. 

For $b>5/2$, the potential has the shape of a plugged volcano; the mountain is still formed by the $\sim 1/z^2$ bulk potential, but there is a $\delta$ barrier at the origin. The zero mode lives in the $\delta$ well at $z=z_*$. In fact it is clear from glancing at the potential that all the modes should be localized near the IR brane, as is clearly depicted in Fig.\ref{fig2}e. The distinguishing feature of the potential for $b>5/2$ is that there is no potential well at the origin, which is different from the case $b<5/2$.

The dual theory is the same as discussed in the previous section for $b>2$. The interpretation of massless mode is a predominantly composite CFT state, as can be gathered from the low energy expansion of the two-point function (\ref{IRcorr}). At high energies, the leading nonanalytic piece in (\ref{UVcorr}) is the conformal two-point function $ \langle {\cal O}{\cal O}\rangle_{CFT}$. From it we can extract the dimension of ${\cal O}$: dim ${\cal O}= b$. The dimension of the dual CFT operator is symmetric about $b=2$, which can be seen clearly by writing dim ${\cal O}=|b-2|+2$, valid for all values of $b$. The bulk potential (\ref{pot1}) is also symmetric about $b=2$, which is illustrated in Fig. \ref{fig2}.  

The holographic Lagrangian can be defined below the UV scale $\Lambda < k$\cite{lectures}:
\begin{equation}
      {\cal L}_{4D} = -{\widetilde Z}_0(\partial\varphi^0)^2 + m_0^2 (\varphi^0)^2+ 
    \chi \Lambda^{3-b} \varphi^0 {\cal O} +{\cal L}_{CFT}~,
    \label{L+}
\end{equation}
where the couplings ${\widetilde Z}_0, \chi,$ and $m_0,$ follow straightforwardly from inspecting the correlator (\ref{UVcorr}). Examining the interaction term, we see that for $b<3$ the coupling is relevant and hence, strong mixing between the source and CFT sectors produces the mass eigenstates in the theory. Past this point $b>3$ the mixing can be neglected and the massless particle is a CFT bound state.

Although there is some subtlety in the region $5/2>b>3$ where strong mixing occurs, qualitatively we can say that when a particle is subject to the plugged volcano quantum mechanical potential, the holographic description is a massless particle which is a bound state of the CFT interacting with a very massive elementary source field through the operator ${\cal O}$.

\section{Fermion holography}
Bulk fermions living in a slice of AdS$_5$ can also be given a 4D dual description in terms of an elementary source field coupled to a strongly interacting CFT. The holography associated with fermions is richer than that of scalars and very relevant for phenomenology. The holographic interpretation of bulk fermions is thoroughly analyzed in \cite{ferm}. Here we begin with a brief review of their results and then analyze the analogue 1D quantum mechanics problem in parallel with the holography. The Schr\"odinger picture is consistent with the dictionary and proves to be a very useful guide to the holographic interpretation. 

The 5D action for a free massive fermion is\footnote{We will use the mostly minus metric ($\eta_{\mu\nu}={\rm diag}(+,-,-,-,-)$) in this section as in \cite{ferm}.}
\begin{equation}
S=\int d^4x dz \sqrt{g}\Big[\frac{i}{2}\overline{\Psi}\Gamma^M D_M\Psi-\frac{i}{2}\overline{D_M \Psi}\Gamma^M \Psi- c k ~ {\rm sgn}(z) \bar{\Psi}\Psi\Big].\label{actionf}
\end{equation}
In warped space, the gamma matrices are $\Gamma^M=e_A^M\gamma^A$, with $e_A^M$ the funfbien and $\gamma^A=(\gamma^a,-i\gamma^5)$ the familiar Dirac gamma matrices.  The covariant derivative is given by $D_M=\partial_M+\omega_M$, where $\omega_M$ is the spin connection. The dimensionless constant $c$ characterizes the fermion mass in units of the curvature $k$. To be consistent with orbifold symmetry the fermion mass must have odd parity.

The Dirac spinor can be broken up into right and left handed components $\Psi = \Psi_+ +\Psi_-$ by defining $\gamma^5\Psi_\pm=\pm\Psi_\pm$. We must choose one of the fields $\Psi_\pm$ as our source field, but not both. This is due to the fact that the Dirac equation is a first order differential equation, allowing us to specify only one of the boundary values on the UV brane\footnote{This is very interesting because it implies that there are two holographic descriptions of a single bulk theory.}. We follow \cite{ferm} and take $\Psi_-$ as our source field which will allow for an easy comparison of the Schr\"odinger picture and the fermion holography\footnote{In the notation of \cite{ferm}, $\Psi_-=\Psi_L$ and $\Psi_+=\Psi_R$.}. We note that there is a simple correspondence between the dual theories with left and right handed sources, which is discussed in \cite{ferm}.

According to the holographic procedure, the boundary value $\Psi_-^0(x)$ sources an operator ${\cal O}_+$ in the dual theory. Because $\Psi_+$ is free to vary on the UV brane, an additional UV boundary term is required by the variational principle to enforce $\delta S=0$, and it is this boundary term which will contribute to the effective action. By integrating out the bulk, we can compute the self energy $\Sigma(p)$ by taking two functional derivatives of the effective gravity action. We then expand $\Sigma(p)$ in different energy ranges and interpret the dual theory. The 4D description can differ dramatically depending on the value of the fermion mass parameter $c$ as well as the orbifold parity assignments of the bulk fields $\Psi_\pm$. Complementary to this, the value of $c$ and the parity also determine the shape of the quantum mechanical potentials for $\Psi_\pm$. The features of these potentials, such as the location of the $\delta$ wells and barriers, can help us gain intuition about the dual theory.  

\subsection{Schr\"odinger potential}

Expanding the fermion fields as
\begin{equation}
\Psi_\pm(x,z)=\sum_{n=0}^{\infty} \psi^n_\pm(x)h^n_\pm (z),
\end{equation}
we can derive from the action (\ref{actionf}) a nonrelativistic Schr\"odinger equation for the eigenfunctions $h^{n}_\pm(z)$
\begin{equation}
\biggl(-\frac{1}{2}\partial_z^2+V_\pm(z) \biggr)h^{n}_{\pm}=\frac{m_n^2}{2}h^{n}_{\pm},
\label{fermeom}
\end{equation}
where the corresponding quantum mechanical potential $V(z)$ for the fermions is
\begin{equation}
V_{\pm}(z)=\frac{1}{2}c(c\mp1)k^2 A^2(z) \pm c~kA(z)\left[\delta(z)-\delta(z-z_*)\right].
\label{pot2}
\end{equation}
The general solution is given by
\begin{equation}
    h_\pm^n(z) =
          N_n A^{-1/2}(z) \biggl[J_{c\mp1/2}\biggl(\frac{m_n}{k A(z)}\biggr)+\beta_n Y_{c\mp1/2}\biggl(\frac{m_n}{k A(z)} \biggr)\biggr],
\end{equation}
where again $N_n$ and $\beta_n$ are determined by boundary conditions and normalization. Unlike the scalar case the solutions of the Schr\"odinger problem $h_\pm^n$ are equivalent
to the physical propagating modes in the bulk. We note that in a supersymmetric extension, the scalar and fermion masses are related by $b=3/2\pm c$ \cite{gp1}. In this case the Schr\"odinger potentials (\ref{pot1}) and (\ref{pot2}) are identical.

The potentials have the same general shapes as in the scalar case depicted in Fig. \ref{fig2}. Consider as an example the potential $V_-(z)$ for different values of $c$. For $c>0$, the fermion propagates in the volcano potential. The potential becomes negative in the region $-1<c<0$, in particular taking the minimum possible value at $c=-1/2$ (analogous to Fig. \ref{fig2}c). As in the scalar case, we will see that this corresponds to the crossover point in the dual interpretation. For $c<-1$, the potential assumes the plugged volcano shape. The potential $V_+(z)$ takes on similar shapes, but for different values of $c$.

The shape of the ``source'' potential $V_-(z)$ can hint as to where the strong mixing occurs. Like the scalar, when the bulk potential is positive, we can roughly say that the mixing between the source and CFT states is negligible. In these realms, the 5D features of the theory can be a very useful guide to the dual description. Localization of the massless mode and location of the $\delta$ wells are reflected in the holographic interpretation. However when the bulk potential $V_-(z)$ becomes negative, there is strong mixing and we really must rely on the correlator $\Sigma(p)$ to interpret the dual theory. There are some subtleties, as in the scalar case, in that the point where the potential becomes negative does not precisely correspond to the point when the source-CFT coupling becomes marginal, but the statement is qualitatively accurate.  
 
In orbifold theories, we must assign parities to the fields at each boundary. If we assign a field to be odd at a particular boundary, this corresponds to assigning a Dirichlet condition. This can be derived by integrating the Schr\"odinger equation (\ref{fermeom}) and assuming that the derivative of the wavefunction, $\partial_z h^{n}_{\pm}$, is continuous across the boundary. This is different from what is ordinarily done in the analogue quantum mechanics problem in which Dirichlet conditions are forced by putting a hard wall at the boundary. Conversely, even parity is accomplished by assuming that the derivative is discontinuous across the boundary, which is the standard way of solving the analogue delta-function quantum mechanics problem\footnote{See Griffiths\cite{griffiths} for a discussion of boundary conditions in the delta-function potential problem.}. Of course, we have additional symmetries in the 5D orbifold theory that forces one of the fields $\Psi_\pm$ to have odd parity. 

To keep in mind which boundary conditions are being imposed, it is useful to represent a hard wall at the boundary if we are imposing Dirichlet conditions, while retaining the $\delta$-function if we are choosing Neumann conditions. 
This provides a pictorial which is helpful when trying to grasp the properties of the modes, such as whether or not there is a zero mode and where such a mode is localized. Thus, for even parity fields, we will retain the delta-function piece, while for odd parity fields, we will simply put a hard wall at the boundary. These are in fact the correct analogue quantum mechanical potentials in the region from $z=0$ to $z=\pi R$ \footnote{This discussion also applies to the scalar if we choose different parity conditions. We didn't mention it in the previous section because we were always interested in keeping a massless mode, which corresponds to even parity.}. This will allow us to differentiate between the potentials $V_+(z)$ and $V_-(z)$ and help us understand the holographic theory. 

For example, if we choose $\Psi_-$ to have even parity $(+ +)$ on each brane, then $\Psi_+$ necessarily will have odd parity $(- -)$. This corresponds to keeping the $\delta$ potentials in $V_-(z)$ and putting a hard wall at each boundary for $V_+(z)$, as shown in Fig. \ref{fig3}a. In the Kaluza-Klein picture, we know that there will be a chiral zero mode $\Psi^0_-(x)$, while all modes $\Psi^n_+$ will be massive. In fact the holography in this case is very similar to the scalar case discussed previously, and there are three important regions with different interpretations. When $c>0$, there is a volcano potential, and the dual interpretation of the massless mode is a primarily source field. In the intermediate region $-1<c<0$, there the massless particle arises out of strong mixing between the source and CFT sectors, and the potential becomes negative. For $c<-1$ the potential is shaped like a plugged volcano, shown in Fig. \ref{fig4}b, and the chiral fermion in the dual theory is predominantly a CFT bound state.   

Of course there are different parity conditions we can choose for the fermions $\Psi_\pm$, and the dual interpretation depends on these choices. The Schr\"odinger potential is also a very useful guide to the dual description in these cases. We will illustrate this with several examples, each for different regions of the boundary mass parameter $c$.

\subsubsection{$c>0$}

We first focus on the case $c>0$ in which $V_-(z)$, the ``source'' potential, is always positive in the bulk. In this region, the mass eigenbasis (KK modes) is approximately the same as the holographic basis (source and CFT fields). In other words, the mixing between the source and CFT are negligible. The potential for each possible choice of parity assignments for $\Psi_-$ and $\Psi_+$ will be analyzed and compared to the dual theory.

\begin{figure}
\centerline{
\includegraphics[width=1.1\textwidth]{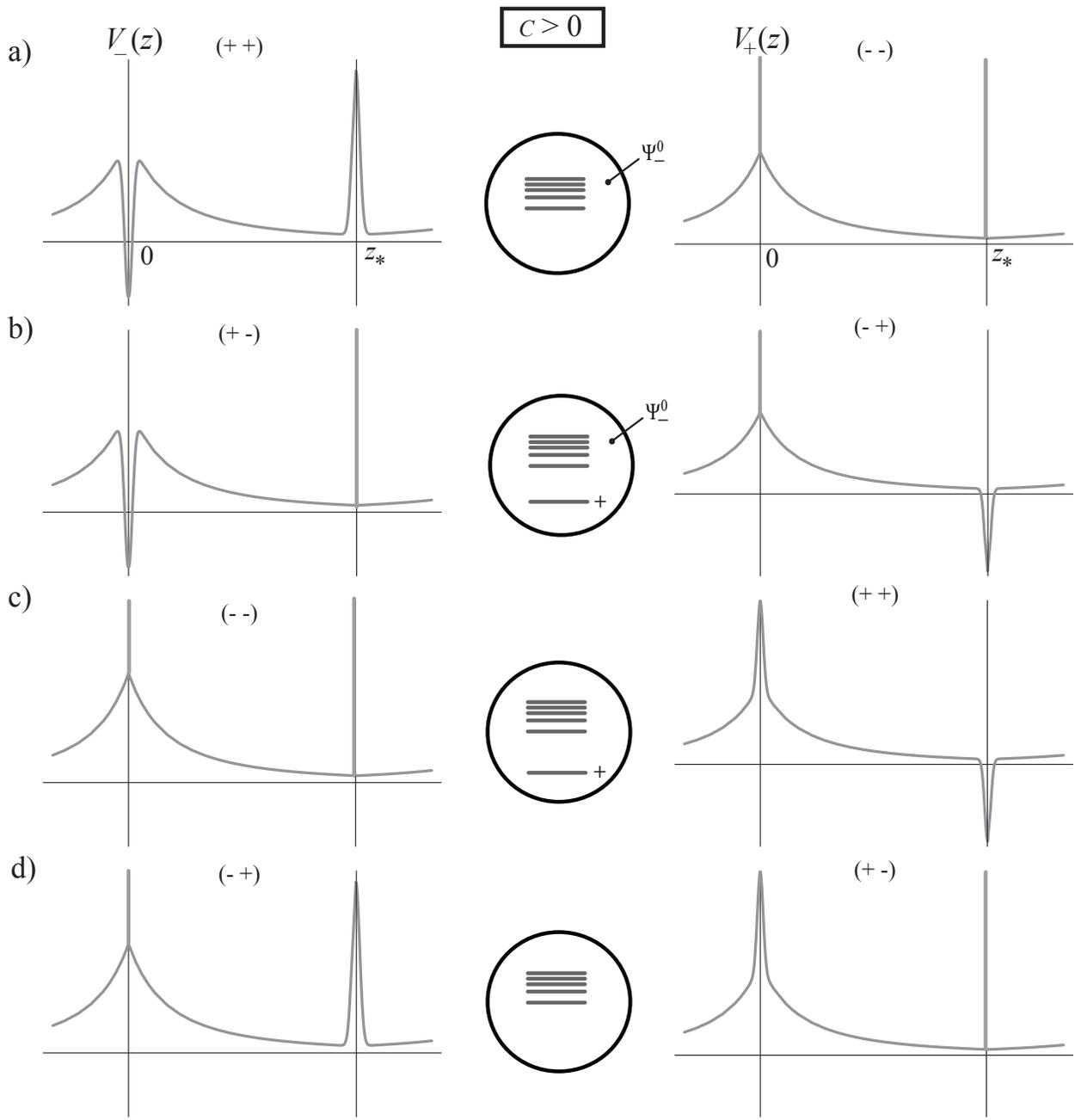}}
\caption{Schr\"odinger potential and dual description for $c>0$ for different orbifold parities. The ``source'' potential $V_-(z)$ is shown on the left while $V_+(z)$ is shown on the right.}
\label{fig3}
\end{figure}

Consider first assigning $\Psi_-$ even parity at each brane $(+ +)$ and $\Psi_+$ odd parity $(- -)$. The situation is depicted in Fig. \ref{fig3}a. The potential $V_-(z)$ has $\delta$-function potentials at each brane, and takes the shape of the volcano near the origin. The $\delta$ well on the UV brane supports a single bound state, which is the massless mode. Thus the massless mode is localized near the UV. On the other hand, the potential $V_+(z)$ simply has a hard wall on each boundary, representing the Dirichlet BC for $\Psi_+$. Only massive modes propagate in this well. 

Let us use the intuition we have gained by studying the scalar case to guess the dual description. Since there is a $\delta$ well at the origin, and thus a massless mode localized on the UV brane, we expect that this is mirrored in the dual theory as a massless source field $\Psi^0_-$. This can be checked by expanding the correlator $\Sigma(p)$, in which case we find the leading analytic term is interpreted as a kinetic term for the source \cite{ferm}. We do not find a pole in the low energy expansion indicating that the CFT is not chiral, as we expect since neither $V_-(z)$ or $V_+(z)$ have a $\delta$ well on the IR brane.

What if $\Psi_-$ has odd parity at each boundary $(- -)$ and $\Psi_+$ has even parity $(+ +)$? In this case the Schr\"odinger potential $V_+(z)$ has $\delta$-function potentials at the boundaries, while instead $V_-(z)$ has hard walls, as in Fig. \ref{fig3}b. In the Kaluza-Klein picture, now $\Psi_+$ contains a massless mode which is localized near the IR brane, since the $\delta$ well is located at $z=z_*$. There are only massive $\Psi_-$ modes. 

In the dual theory, we would wager that the CFT is chiral, since there is a $\delta$ well on the IR brane. Indeed we do find a massless pole in the low energy expansion in this case. What about the source field? Although we find a kinetic term in the expansion of $\Sigma(p)$, the source field is actually frozen in the dual theory. To impose the Dirichlet condition on the UV brane we add a term containing a lagrange multiplier to the UV Lagrangian which enforces the boundary condition \cite{ferm}. This term, when included in the effective action, implies that the source field is not dynamical in the dual theory; in other words, $\Psi^0_-$ is simply a classical source for the operator ${\cal O}_R$. We can always tell from the pictorial whether or not the source is dynamical by examining whether or not it has a $\delta$-function potential at the UV boundary.

Next we consider mixed boundary conditions, with $\Psi_-$ having $(+ -)$ parity and $\Psi_+$ having $(- +)$. The potentials are shown in Fig. \ref{fig3}b, with $V_-(z)$ having a $\delta$ barrier at the origin and $V_+(z)$ having a $\delta$ well at $z=z_*$. We know that there are only massive modes when the fermions have opposite parity. The only way a massless mode could occur is if the potential has delta-functions at {\it each} boundary, which is not the case here. However, we expect that the source field propagates in the dual theory because it has been assigned an even UV parity, and further, we would guess that the CFT is chiral by the presence of the $\delta$ well located at the IR boundary. In fact, we do find a pole in the low energy expansion of the correlator \cite{ferm}, and the source field does propagate in the dual theory (no lagrange multiplier term exists in the UV action). These two states are connected through a mass mixing in such a way that there are no massless eigenstates in the dual theory. Once again, the Schr\"odinger picture appears consistent with the dual theory.

Finally, let us examine the case with $\Psi_-$ having $(- +)$ parity and $\Psi_+$ having $(+ -)$. Again, we know that there are no massless modes in the 5D theory. The difference here from the previous mixed BC case is that we do not expect a propagating source because there is a hard wall on the UV boundary in $V_-(z)$, and we do not expect the CFT to be chiral because the $\delta$ well is absent on the IR boundary in $V_+(z)$, illustrated in Fig \ref{fig3}d. Our expectations are confirmed in the dual theory, which contains only massive CFT states.

\subsubsection{$-1<c<0$}

In this region the potential $V_-(z)$ is negative in the bulk, and this corresponds to strong mixing in the CFT. We cannot solely rely on 5D properties such as localization features or location of $\delta$ wells in the potential as a guide to holography, unlike the case when the mixing is negligible. Instead we must carefully examine the correlator to make the dual interpretation, but at least the potential gives us warning as it turns negative.

However, we can still glean some useful information from the potential. At the point $c=-1/2$ the bulk potential $V_-(z)$ reaches its minimum value. From our experience with the scalar, we guess that this point precisely corresponds to the crossover point in the dual interpretation, and this is in fact the case. Consider the situation in which the field $\Psi_-$ is assigned even parity on each boundary $(+ +)$ while $\Psi_+$ is assigned odd parity $(- -)$. As we discussed previously, for $c>-1/2$ the massless particle in the dual theory is mostly source field. However, for $c<-1/2$, the composition of the massless eigenstate is dominated by a CFT bound state \cite{ferm}. These descriptions, as always, are interpreted from expanding the correlator computed with holographic recipe. However, the minimum of the Schr\"odinger potential indeed corresponds to the transition point in the holographic interpretation of the bulk theory. Moreover, the bulk potential $V_-(z)$ is symmetric about the transition point, which is replicated in the dimension of the dual operator dim~${\cal O}_+=3/2+|c+1/2|$.

\subsubsection{$c<-1$}

Finally we consider the case $c<-1$, in which the potentials are again always positive in the bulk. In particular for $c<-3/2$ the mixing is negligible and the mass eigenbasis is approximately the same as the source and CFT fields. In this region, we can trust the Schr\"odinger potential as an effective guide to the holographic description. 

\begin{figure}
\centerline{
\includegraphics[width=1.1\textwidth]{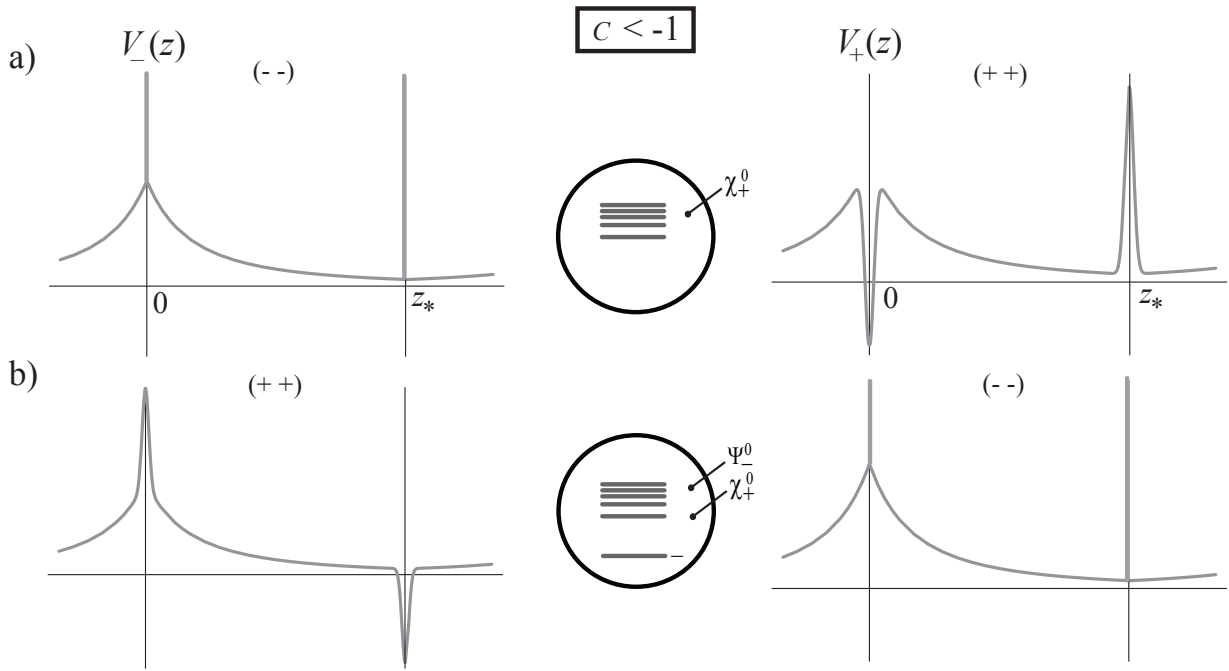}}
\caption{Schr\"odinger potential and dual description for $c<-1$ for different orbifold parities. The ``source'' potential $V_-(z)$ is shown on the left while $V_+(z)$ is shown on the right.}
\label{fig4}
\end{figure}

A new feature appears in this region that we did not encounter with the scalar case. Let us assign odd $(- -)$ parity to $\Psi_-$ and even $(+ +)$ to $\Psi_+$. The potentials are drawn in Fig. \ref{fig4}a. Hard walls indicate the Dirichlet conditions for $\Psi_-$, and we therefore know that only massive modes live in $V_-(z)$. However, $V_+(z)$ contains $\delta$ potentials at the boundaries, in particular a $\delta$ well at the origin. We know a massless mode is bound in this well. 

What do we expect the holographic interpretation to be? We know that the source must be frozen because of the odd UV parity. However we also see that neither potential has a $\delta$ well on the IR boundary, indicating that the massless particle is not a CFT state. The delta well on the UV brane in $V_+(z)$ would seem to indicate that the holographic description would include a new elementary degree of freedom $\chi_+$ not associated with the source field. Indeed this is the case. We find that even in the high energy expansion of $\Sigma(p)$ a pole appears \cite{ferm}. This pole can not be attributed to the CFT because conformal symmetry is only mildly broken in this energy regime. The next to leading nonanalytic term is interpreted as the two point function for the CFT operator ${\cal O}_+$, and therefore we must interpret this additional pole as arising from the pointlike $\chi_+$, which is external to the CFT. We see that the Schr\"odinger potential is indeed consistent with the dual theory.

If we instead assign even $(+ +)$ parity to $\Psi_-$, the $\delta$ well on the IR brane in $V_-(z)$ (Fig. \ref{fig4}b) indicates that the massless mode is a chiral CFT bound state. The source field propagates and mixes with the elementary $\chi_+$ to form a massive state. Other parities can also be analyzed for $c<-1$ and the potential again serves as a helpful tool in analyzing the dual interpretation.  

\section{Examples from phenomenological models}
One of the more interesting and relevant examples from electroweak phenomenology that we can discuss from the Schr\"odinger point of view is the standard model in the bulk, where zero mode standard model fermions are localized at different points in the warped dimension. This nicely explains the fermion mass hierarchy without any flavor problems from the bulk  fermions~\cite{gp1, hs}. There are also interesting deviations from standard model flavor physics that will be seen at the LHC~\cite{huber,aps}. The two most extreme examples are the electron and the top quark. The electron has a tiny Yukawa coupling and thus is localized on the UV brane so that it weakly overlaps with the Higgs on the IR brane. Conversely, the top quark is localized on the IR brane to produce a large Yukawa coupling. 

The profile of the fermion fields, and thus their effective Yukawa couplings depends on the value of the fermion mass $c$. For concreteness, let us take $\Psi_-$ to have even parity and therefore a localized zero mode. Then for the electron, the coupling is roughly $\lambda_e\sim 10^{-6}$, which corresponds to a mass of $c\sim 0.64$. Thus, the electron lives in the volcano potential $V_-(z)$ shown in Fig. \ref{fig3}a. In the dual interpretation, it is clear that the massless particle is mostly an elementary source field. The top quark, on the other hand, will have a mass $c\sim -0.5$ to produce the $y_t\sim1$ Yukawa coupling. This has a negative potential in the bulk, corresponding to the special symmetric point of a partly elementary, partly composite dual top quark. 

Gauge fields propagating in the bulk have also been studied extensively in the literature \cite{gaugeb1,gaugeb2}. Because the zero mode is not localized, but rather flat in the extra dimension, it is often stated that the corresponding massless field in the dual theory is a mixture of composite and elementary states. As we saw earlier, absolute localization in the fifth dimension is not always the best guide to the nature of the dual theory, and in fact, the massless particle in the 4D theory is primarily an elementary state external to the CFT. This can be understood from the Schr\"odinger potential. If we expand the gauge field as $A_\mu(x,y)=\sum_{n=0}^\infty A_\mu^n(x) A(z)^{-1/2}g^n(z)$, we can derive the Schr\"odinger equation of motion for $g^n(z)$:
 \begin{equation}
\biggl(-\frac{1}{2}\partial^2_z+
\frac{3}{8}k^2 A^2(z)-\frac{1}{2}k A(z)\biggl[\delta(z)-\delta(z-z_*)\biggr]\biggr)g^n(z)=\frac{m_n^2}{2}g^n(z)~.
\end{equation}
We see that the modes propagate in a volcano potential, indicating that the zero mode indeed corresponds to an elementary source field in the dual theory. This can be confirmed by analyzing the gauge field two-point function \cite{pheno1, ad, u1}. 

Perhaps an even more striking phenomenological example is provided by models that identify the Higgs boson as the scalar component of a higher dimensional gauge field~\cite{cnp,acp}. The equation of motion for the zero-mode $A^0_5$ can be written in the form of a Schr\"odinger equation: 
\begin{equation}
\biggl(-\frac{1}{2}\partial^2_z-\frac{1}{8}k^2 A^2(z)+\frac{1}{2}k A(z)\biggl[\delta(z)-\delta(z-z_*)\biggr]\biggr)A_5^0(z)=0~.
\end{equation}
In fact this is simply the equation for the scalar field (\ref{schr1}) with $b=2$. The potential, shown in 
Fig. \ref{fig2}c, is at its minimum value with respect to $b$, corresponding to the transition point in the dual interpretation. Thus, we would say that the dual Higgs field is, in some sense, ``equally'' composed of composite CFT fields and source field. Interestingly, the 5D gauge symmetry forces the $A_5$ field to have the particular potential with $b=2$. This is different from the fermion case, where the mass parameter is arbitrary and can take a range of phenomenologically viable values. 

\section{Conclusion}
We have derived the Schr\"odinger potentials for localized modes in warped extra dimensions. This gives a clear and simple picture of localization in the extra dimension. The Schr\"odinger potentials were also compared to the known AdS/CFT dictionary relating gravity in a slice of AdS$_5$ and a strongly coupled 4D CFT. The Schr\"odinger potential assumes different shapes depending on the mass parameters of the theory, corresponding to different 4D dual theories, and thus provides a useful qualitative tool for holography. At the transition point in the dual interpretation, the bulk potential takes on its minimum value.  In particular, for bulk scalar fields the bulk potential is symmetric about $b=2$, which mimics a similar symmetry in the dimensions of operators in the dual theory. This transition point in the dual theory is different from that expected merely from the localization of the flat-metric zero mode (which has a symmetry at $b=1$). The dual picture for bulk fermions is more involved and subtle, but by using the Schr\"odinger potential it becomes more apparent. It makes clear when extra elementary states are needed in order to obtain a consistent holographic interpretation and displays the symmetry in the operator dimension about $c=\pm 1/2$. Finally, our analysis suggests that perhaps other Schr\"odinger potentials with similar structure can be given a dual interpretation, and this could lead to novel applications of the AdS/CFT correspondence.

\section*{Acknowledgments}
The work of B.B. was supported in part by a Department of Energy 
grant DE-FG02-94ER40823 at the University of Minnesota and an award from Research Corporation. The work of A.L. was supported by the National Science Foundation grant NSF/PHY-0139099. We thank Tony Gherghetta for helpful conversations. A.L. acknowledges the University of Minnesota REU program.

\end{document}